\documentclass[%
 reprint,
superscriptaddress,
frontmatterverbose, 
nofootinbib,
 amsmath,amssymb,
 aps,
 showpacs,
pra,
]{revtex4-1}

\usepackage{graphicx}
\usepackage{dcolumn}
\usepackage{bm}

\usepackage{geometry}
\usepackage{verbatim}
\usepackage{mathrsfs}
\usepackage{amsmath}
\usepackage{amssymb}
\usepackage{esint}

\usepackage{tikz}
\usetikzlibrary{shapes.misc}
\usetikzlibrary{decorations.markings}
\usetikzlibrary{calc}

\usepackage[unicode=true,pdfusetitle,
 bookmarks=true,bookmarksnumbered=false,bookmarksopen=false,
 breaklinks=false,
colorlinks=true]
 {hyperref}

\usepackage{soul}

\makeatletter

\pdfpageheight\paperheight
\pdfpagewidth\paperwidth

\providecommand{\LyX}{\texorpdfstring%
  {L\kern-.1667em\lower.25em\hbox{Y}\kern-.125emX\@}
  {LyX}}
\providecommand{\tabularnewline}{\\}

\makeatother

\begin{document}

\def\plotsheight{6cm}

\title{Thermodynamics of Quantum Phase Transitions of a Dirac oscillator in a homogenous magnetic field}

\author{\textsc{A.~M.~Frassino}}
\affiliation{Frankfurt Institute for Advanced Studies, Ruth-Moufang-Stra\ss e 1, D-60438 Frankfurt am Main, Germany}

\author{\textsc{D.~Marinelli}}
\affiliation{Machine Learning and Optimization Lab., RIST,  400487 Cluj-Napoca, Romania}
\email[ Email: ]{marinelli@rist.ro}

\author{\textsc{O.~Panella}} 
\affiliation{Istituto Nazionale di Fisica Nucleare, Sezione di Perugia, Via A.~Pascoli, I-06123 Perugia, Italy}
\email[ Email: ]{orlando.panella@pg.infn.it}

\author{\textsc{P.~Roy}}
\affiliation{Physics and Applied Mathematics Unit, Indian Statistical Institute, Kolkata-700108, India}

\date{\today}

\begin{abstract}
The Dirac oscillator in a homogenous magnetic field  exhibits a chirality phase transition at a particular (critical) value of the magnetic field. Recently, this system has also been shown to be exactly solvable in the context of noncommutative quantum mechanics featuring the interesting phenomenon of re-entrant phase transitions. In this work we provide a detailed study of the thermodynamics of such quantum phase transitions (both in the standard and in the noncommutative case) within the Maxwell-Boltzmann statistics  pointing out that the magnetization has discontinuities at  critical values of the magnetic field even at finite temperatures.
\end{abstract}

\maketitle

\section{Introduction}
\label{sec:introduction}

Quantum Phase Transitions (QPT) \cite{Sachdev201105} are a class of phase transitions that can take place at zero temperature when the quantum fluctuations, required by the Heisenberg's uncertainty principle, cause an abrupt change in the phase of the system.
The QPTs occur at a critical value of some parameters of the system such as pressure or magnetic field. 
In a QPT, the change is driven by the modification of particular couplings that characterise the interactions between the microscopic elements of the system and the dynamics of its phase near the quantum critical point.

For a quantum system at finite temperature $T$, both the thermal and the quantum fluctuations are present. The interplay between the quantum and the thermal
fluctuations can either smooth out the differences between the phases
(namely, no phase transition occurs at finite temperature) or there
can be regimes in which some discontinuities hold, 
and a phase transition appears. 
Eventually, the thermal fluctuations prevail \cite{Sachdev201105}.

In this paper, we focus on a particular quantum system, namely the Dirac oscillator, at finite temperature $T$ and 
with an additional constant magnetic field $\textbf{B}$.
The Dirac oscillator system including an interaction in the form of a homogeneous magnetic field is an exactly solvable system \cite{PhysRevLett.99.123602,2007PhRvA76Delgado,PhysRevA.77.033832,Mandal:2009we,Mandal:2012wp} and shows interesting properties. In particular, at zero temperature, if the magnitude of the magnetic field
either exceeds or is less than a critical value $B_{\text{cr}}$ (which 
depends on the oscillator strength), this combined system shows a
\emph{chirality phase transition}  \cite{2008PhRvA77:Delgado,Quimbay:2013gta}. Because of
the 
phase transition, the energy spectrum is different
for $B>B_{\text{cr}}$ and $B<B_{\text{cr}}$ where $B$ is the magnetic field strength.

\begin{figure}
\begin{centering}
\includegraphics[scale=0.6
]{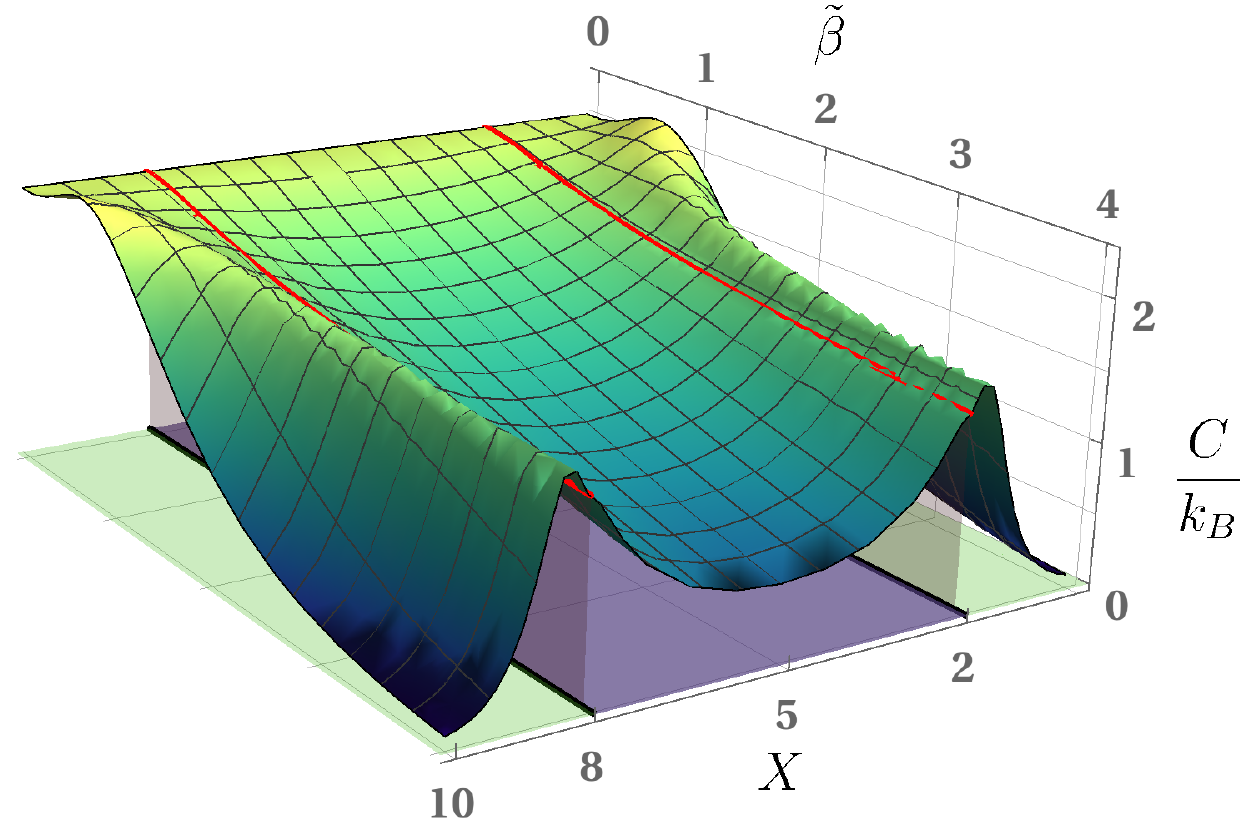} 
\caption{Re-entrant phase transition as shown by the heat capacity of 
a noncommutative Dirac Oscillator  as a function of the  scaled effective magnetic energy  $X\equiv \left(\mu\, B - \hbar\omega\right)/mc^2$ and the  scaled inverse temperature $\tilde{\beta}= mc^2 / k_B T$. 	$X_1={\omega_{\eta}\hbar}/{mc^{2}}=2$, $X_2=\left({\omega_{\theta}\hbar}/{mc^{2}}\right)=8$  are the two critical points that divide the green regions (left phases) from the violet region (right phase). In the left phases, the heat capacity is non-monotonic in $\tilde{\beta}$. }\label{fig:CNNcom}
\end{centering}
\end{figure} 

The $(2+1)$-dimensional Dirac oscillator in the presence of a constant magnetic field has also been studied at zero temperature
in the framework of noncommutative space coordinates and momenta
 \cite{Panella:2014hga}. It has been shown that in this case  $B_{\text{cr}}$ depends not only on the oscillator strength but also on the noncommutative parameters. An 
interesting consequence of the noncommutative scenario is that, apart from the left- and right-chiral
phases of the commutative case \cite{PhysRevA.77.033832}, there is also a third left phase
and thus a second quantum phase transition (right-left) 
\cite{Panella:2014hga} (see Fig. \ref{fig:CNNcom}). The presence of the third phase with left chirality leads to what is called
a Re-entrant Phase Transition (RPT) \cite{2016:PanellaRoyReE} first observed in nicotine/water mixture \cite{hudson1904gegenseitige} as noted in \cite{Mann2016}. 
This phenomenon has been recently found also in gravitational systems \cite{Altamirano:2013ane,Frassino:2014pha}.

In this work, we analyze the Dirac oscillator  in the presence of a constant magnetic field at finite temperature.
While in the first part of the paper we consider the standard system at finite $T$, the second part will be 
dedicated  to the noncommutative Dirac oscillator.

To study what kind of QPT the Dirac oscillator at finite $T$ undergoes, we will 
investigate the system at high temperatures
when the statistics can be approximated with the Maxwell-Boltzmann 
one \cite{Kubo1990}.
We find that in this regime the QPT does not disappear and therefore the quantum fluctuations prevail upon the thermal ones (as can be  seen in the noncommutative case in Fig.~\ref{fig:CNNcom}). Only in the limit of very high temperatures ($\beta \rightarrow0$) the QPT disappears.

The organization of the paper is as follows: in Sec.\ref{sec:System} we
shall present the system to be analyzed and discuss the spectrum of the two phases, namely, the left and the right phase; in Sec. \ref{sec:FiniteT}, we
present the method to calculate the partition function that characterizes the chiral phases at finite temperature and describes the phase transition as the strength of the magnetic field varies; in Sec. \ref{sec:magnetiz} we scrutinize in details the magnetization at finite temperature and in Sec. \ref{sec:NCoscillator} we study the thermodynamic of a noncommutative Dirac Oscillator with a constant magnetic field. Finally, Sec. \ref{sec:conclusions} is devoted to conclusions.

\section{The Quantum System: Dirac Oscillator with a constant magnetic field} \label{sec:System}

The non-relativistic version of the
 $\left(2+1\right)$-dimensional Dirac oscillator with a constant magnetic field and zero temperature can be associated to a chiral harmonic oscillator \cite{2008PhRvA77:Delgado}, that  have been
studied in \cite{Banerjee:1998em}. Moreover, in  \cite{HortaBarreira:1992ec,Dunne:1989hv} a possible connection to topological Chern-Simons gauge theories has been pointed out. The case of the system presented in this paper can be seen as a relativistic extension of the chiral harmonic oscillators \cite{2008PhRvA77:Delgado,Quimbay:2013gta}, at finite temperature.

A relativistic spin-1/2 fermion constrained in a 2-dimensional plane  with mass $m$, 
charge $e$,  Dirac oscillator frequency $\omega$ and subjected to a a constant magnetic field orthogonal to the plane,  is described by the Hamiltonian:
\begin{equation}
H= c\, \bm{ \sigma} \cdot
  (\bm{p}-im\omega\sigma_z \bm{x}+\frac{e}{c}\bm{\,A})
  +\sigma_z \,mc^{2}\,, \label{eq:hamiltonian}
\end{equation}
where $c$ stands for the speed of light
and $\bm{\sigma}=\left(\sigma_x,\sigma_y\right)$, $\sigma_z$
denotes the Pauli matrices. 
As in the usual notation, the $\bm{p}$ and the $\bm{x}$ represent the momentum and the position operators
while the vector potential is related to the magnetic field through
\begin{equation}
  \bm{A}=(-B{y}/2,B{x}/2)\,.
\end{equation} 
This set-up offers an intriguing
interplay: while the two-dimensional Dirac oscillator coupling
endows the particle with an intrinsic left-handed chirality \cite{2007PhRvA76Delgado}, the magnetic field coupling favors
a right-handed chirality \cite{PhysRevLett.99.123602}. The interplay between opposed chirality interactions culminates in the appearance
of a relativistic quantum phase transition, which can be fully characterized \cite{2008PhRvA77:Delgado}.
\subsection{Energy levels}\label{sec:energy-levels}
In this section, we shall recall the spectrum and the degeneracy of
the various energy levels of a two-dimensional relativistic Dirac oscillator in the two phases defined by $B>B_{\text{cr}}$ and $B<B_{\text{cr}}$.
The spectum of the system is characterized by two quantum numbers: $n_r = 0,1,2,...$, the radial quantum number
and $M=0,\pm 1, \pm 2,...$, the 2-dimensional angular momentum quantum number.

In what we call the \emph{left phase}, the energy level corresponding to the zero mode has positive energy $E=\, mc^2$ and infinite degeneracy with respect to the non-negative magnetic quantum number
$M\ge 0$ (the negative magnetic numbers are forbidden \cite{Panella:2014hga}). The excited states are
\begin{eqnarray}
E_{N}^{\pm}=\,\pm mc^{2}\sqrt{1+\,\xi_{L}\,N}\,,\quad\\
N=n_{r}+\frac{|M|-M}{2},\quad N=1,2,\,,\dots\,\label{spectrumL1}
\end{eqnarray}
where $N$ labels the energy levels, $\xi_L$ is a constant encoding  
the parameters of the system  that for the standard commutative case simply reads 
\begin{gather}
   \xi_L = -4\,X, \quad X := \frac{1}{mc^{2}}\left(\mu\, B - \hbar\omega\right)
     \label{eq:xiLcomm}
\end{gather} 
and $\mu=e \hbar / (2 m\, c)$ denotes the Bohr magneton.
The energy levels in \eqref{spectrumL1} are degenerate. 
In particular, every level has infinite degeneracy, with respect  to the non-negative values of $M$, 
and $D=N+1$ finite degeneracy with respect to the negative values of $M$. \\
The \emph{right phase}, on the other hand,  has zero mode degeneracy with respect to the non-positive magnetic
quantum number $M \le 0$ and the value of the energy is in the negative branch: $E = - m\, c^2$.
Similarly, the excited states have  an infinite degeneracy with respect  to non-positive values 
of $M$, while the degeneracy is $D=N+1$ with respect to $M>0$ and the energies read
\begin{multline}
E_{N}^{\pm}=\,\pm mc^{2}\sqrt{1+\,\xi_{R}\,N}\,,\\
N=n_{r}+\frac{|M|+M}{2}\quad N=1,2,\dots\,,\label{spectrumR}
\end{multline}
where the parameter $\xi_{R}$ is related to $\xi_{L}$ by the relation $\xi_{R}=-\xi_{L}$. Note that, when $\xi_{R} = \xi_{L} = 0$, that is when 
\begin{equation}
B = B_{\text{cr}} =\hbar\,  \mu \,  \omega, \label{eq:BcrComm}
\end{equation} 
 all the energy levels collapse to the energy $E=\pm mc^2$ and  the \emph{quantum phase transition} happens \cite{2008PhRvA77:Delgado,Panella:2014hga}.
 
\section{Finite Temperature Quantum Phase Transition} \label{sec:FiniteT}

To study the interplay between thermal and quantum fluctuations for this two-dimensional relativistic Dirac oscillator, we focus on the system at high temperatures when the electron-states statistics  can be described with the Maxwell-Boltzmann statistics. We will show that in this regime the QPT will not disappear.

\subsection{Partition Function}

The partition function for the positive branch of the energy levels (\ref{spectrumL1}) and (\ref{spectrumR}) and the zero mode, has to take into account the degeneracy of each single energy level. Thus, it is defined by the product of the density of the states and the Boltzmann factor
\begin{gather}
Z_{L,R} =  \sum_{M,n_r}\, \mathtt{g} \, e^{-\beta\, E^{+}_{N}}, \quad N=0,1,\dots\,, \label{eq:Z-LR}
\end{gather}
where $E^{+}_{N}$ are the values given by (\ref{spectrumL1}) and (\ref{spectrumR}), and $\beta$ is the inverse temperature.
\begin{figure}[]
	\begin{center}

	    \begin{tikzpicture}[scale=2]

	    \def\xz{0}
	    \def\xs{1}
	    \def\xzs{1.8}
	    \def\xss{3.5}

	    \def\DiscLev{$+m\,c^2$/E0/0,
	    	$+m\,c^2 \sqrt{1+\xi_{R,L}}$/E1/0.5,
	    	$+m\,c^2 \sqrt{1+2\,\xi_{R,L}}$/E2/1,
	    	$+m\,c^2 \sqrt{1+N\,\xi_{R,L}}$/E4/1.8}

	    \def\ConStart{0,   0.25, 0.75,1.25,1.75}
	    \def\ConStop {0.25,0.75, 1.25,1.75,2.25}

	    \def\Fromto
	    {0.25/0.75/0.5,0.75/1.25/1.0,1.55/2.05/1.8}
	    
	    \foreach \form/\lab/\i in \DiscLev {
	    	\draw (\xzs,\i)   to node [pos=0.6,above=-3pt,font=\small] (\lab){\form} (\xss,\i) ;
	    }
	    

	    \foreach \from/\to/\mid in \Fromto 
	    \foreach \k in {0,0.05,...,0.5} {
	    	\draw ($(\xz,\from)+(0,\k)$)  to  ($(\xs,\from)+(0,\k)$) ;
	    	\draw  ($(\xs,\from)+(0,\k)$) to[out=0,in=180]  (\xzs,\mid) ;
	    	
	    }
	    
	    \foreach \k in {0,0.05,...,0.25} {
	    	\draw[red] ($(\xz,0)+(0,\k)$)  to  ($(\xs,0)+(0,\k)$) ;
	    	\draw[red]  ($(\xs,0)+(0,\k)$) to[out=0,in=180]  (\xzs,0) ;
	    	
	    }
	    
	    \draw[dotted,thick, blue] (0.5,1.28) --   (0.5,1.50) ;
	    \draw[dotted,thick, blue] (3.0,1.3) --   (3.0,1.7) ;
	    
 	    \draw[dotted,thick, blue] (0.5,2.1) --   (0.5,2.250) ;
 	    \draw[dotted,thick, blue] (3.0,2.1) --   (3.0,2.25) ;
	    
	    
	    
	    \def\xarrow{-0.2}
	    
	    \draw[->,thick,gray] (\xarrow,0) to 
	     node[above,sloped,font=\scriptsize,pos=0.8] {$mc^2\sqrt{1+p^2}$} 
	     node[above,sloped,font=\scriptsize,pos=0.2] {$\omega=B=0$}  (\xarrow,2.255)  ;

	    \foreach \i in \ConStart{
	    	\foreach\j in \ConStop{
	    		
	    	}
	    }
	    
	    \end{tikzpicture}
   	\end{center}
   	\caption{Pictorial view of the positive branch of the  energy levels that from a continuous spectrum for the free electron collapse to the discrete one in \eqref{spectrumL1} and  \eqref{spectrumR}.
   	\label{fig:spectrum} }
\end{figure}
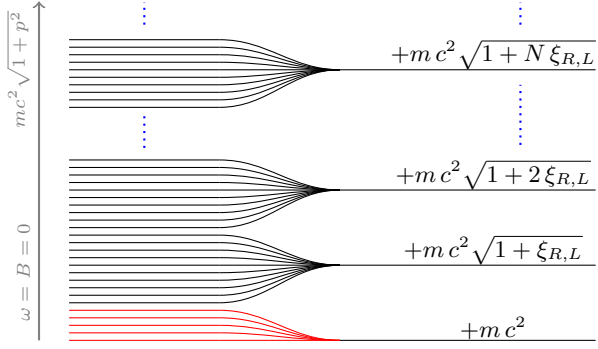

To evaluate the  density of the states $ \mathtt{g}$,
we can split every chiral phase of the system (left and right introduced in the previous section) into two different sub-systems with respect to the sign of the quantum number $M$.

If we consider a free electron with $M>0$ or alternatively $M<0$, in a standard commutative spacetime and
confined to a finite area $\mathtt{L}^{2}$, the number of energy levels in the 
region $dp_{x}dp_{y}$ around $p_{x}$ and $p_{y}$ is $ d\mathtt{g} = \mathtt{L}^{2}/\left( 2 h^{2} \right)dp_{x}dp_{y}$ \cite{Kubo1990}.
Note that it can be proved that the phase-space volume element 
$dp_{x}\wedge dp_{y}$
in the relativistic case is invariant for Lorentz transformations \cite{hakim2011introduction}. Moreover, $\mathtt{L}$ is the length in a rest-frame.

\paragraph{Left Phase.} 
For $M\ge0$, the degeneracy of each energy level (\ref{spectrumL1}) defined in the region $\mathscr{K}$ of the momenta $(p_x,p_y)$ such that $E_N<E_{\text{free}}(p_x,p_y)<E_{N+1}$ (see Fig.\ref{fig:spectrum}), where $E_{\text{free}}$  is the energy of the free electron, is given by
\begin{eqnarray}
  \mathtt{g}_{M>0}&=& 
  \frac{1}{2}\left(\frac{\mathtt{L}}{h}\right)^{2}
  \iintop_{\mathscr{K}}dp_{x}dp_{y} \nonumber \\ &=&
  \pi\left(\frac{\mathtt{L}}{h}\right)^{2}\left[p^{2}\right]_{\frac{p^{2}}{m^{2}c^{2}}=\,\xi_{L}N}^{\frac{p^{2}}{m^{2}c^{2}}=\,\xi_{L}\left(N+1\right)}
  = \left(\frac{\mathtt{L}mc}{h}\right)^{2}\,\pi\xi_{L} .
\end{eqnarray}
and $N$ depends only on the quantum number $n_{r}$ because the difference in \eqref{spectrumL1} is zero.
In the case $M<0$, the number of levels collapsing into the level $N$ is still equal to $\mathtt{g}_{M>0}$. However, one should consider that, when $M<0$, as shown in (\ref{spectrumL1}),
 $N$ depends on both the quantum number $M$
and $n_{r}$, namely  $N = n_r + M $.  Therefore, one needs to keep into account, for each energy level $N$, the degeneracy $D$,  i.e.  $ \mathtt{g}_{M<0}= \mathtt{g}_{M>0}/D$.  
Thus, the number of states 
\begin{equation}
  \mathtt{g}_{M<0}= \left(\frac{\mathtt{L}mc}{h}\right)^{2}\,\frac{\pi\xi_{L}}{\left(n_{r}-M+1\right)},\label{eq:StateDensity}
\end{equation}
 is not constant along the spectrum, but depends on the energy level.
 
 Taking into account  the degeneracy,
 using the 
 spectrum (\ref{spectrumL1}) and its zero point energy, the partition function with Boltzmann statistics of the states of a  single oscillator\footnote{The classical multi-oscillator states partition function is simply $ (Z_L)^n / n! $ for $n$ electrons \cite{Kubo1990}. } for the left phase
is
\begin{eqnarray}
  Z_L &=&
  	\left(\frac{\mathtt{L}mc}{h}\right)^{2}\,\pi\xi_{L}
  \left[
  	e^{-\beta mc^{2}}+
  	\sum_{n_{r}=1}^{\infty}
  	\ e^{-\beta mc^{2}\sqrt{1+\,\xi_{L}\,n_{r}}}+\right.\nonumber\\
  	&+&\left.
	\sum_{n_{r}=1}^{\infty}
 	\sum_{M=0}^{-\infty}
    \frac{e^{-\beta mc^{2}\sqrt{1+\,\xi_{L}\,\left(n_{r}-M\right)}}}
  	{\left(n_{r}-M+1\right)}
  \right]. \label{eq:ZSingleOscillator}
\end{eqnarray}
Notice that the contribution to the partition function of the zero mode energy appears only for $M\ge 0$ \cite{Panella:2014hga} and, as we will see later in sec. \ref{sec:CriticalPoint}, this will be the source of the non-analiticty of the partition function at the phase point.
The sum in the third term of \eqref{eq:ZSingleOscillator} coincides with the second if we introduce $M'$ and $N'$ such that $M'=-M$ then $n_{r}+M'=N'$, obtaining 
\begin{equation}
\label{doublesum}	\sum_{n_{r}=1}^{\infty}\sum_{N'=n_{r}}^{\infty}
    \frac{e^{-\beta mc^{2}\sqrt{1+\,\xi_{L}\,\left(N'\right)}}}			{\left(N'+1\right)}
    =\sum_{N'=1}^{\infty}e^{-\beta mc^{2}\sqrt{1+\,\xi_{L}\,\left(N'\right)}}, 
\end{equation} 
so, over all, the two sectors ($M>0$ and $M<0$) equally contribute to the partition function.

\paragraph{Right Phase}
The partition function for the right phase, defined by the energy levels (\ref{spectrumR}), is equal to (\ref{eq:ZSingleOscillator}) except for the zero mode term. This is because, as explained in Sec. \ref{sec:System}, the zero mode contributes only to the negative energies branch.

\subsection{Zeta-function representation \label{sub:mellin}}
\begin{figure}

\begin{centering}
\begin{tabular}{cc}
 
 \begin{tikzpicture}[scale= 0.28]

\tikzset{->-/.style={decoration={
  markings,
  mark=at position #1 with {\arrow{>}}},postaction={decorate}}}

      \draw[->] (-5,0) -- (5,0) node[right] {$\mathfrak{Re}\left(x\right)$};
    	\draw[->] (0,-5) -- (0,5) node[above] {$\mathfrak{Im}\left(x\right)$};
    	
		  \draw[->-=0.7,blue] (1.5,-5) -- (1.5,5) node[right] {};
		  
		  \node [fill=red,circle, inner sep=1pt,label={[red]-30:$\mathfrak{c}$}]  at  (1.5,0) {};

    	\foreach \s in {0,...,5}
    	{%
    		\draw ( {-\s} , 0 ) node {$\times$};
    	}	
 \end{tikzpicture}
 
 &
 
 \begin{tikzpicture}[scale=0.28]

\tikzset{->-/.style={decoration={
  markings,
  mark=at position #1 with {\arrow{>}}},postaction={decorate}}}

      \draw[->] (-5,0) -- (5,0) node[right] {$\mathfrak{Re}\left(s\right)$};
    	\draw[->] (0,-5) -- (0,5) node[above] {$\mathfrak{Im}\left(s\right)$};
    	
	      \def\rad{5.2}
		  \draw[->-=0.7,blue] (2.5,-\rad) to node[above,sloped,font=\tiny,pos=0.75] {$R\rightarrow\infty$} (2.5,\rad);
		  \draw[->-=0.7,blue] (2.5,\rad) arc (90:270:\rad) node[right,font=\tiny] {};

		  \node [fill=red,circle, inner sep=1pt,label={[red]-30:$\mathfrak{c}$}]  at  (2.5,0) {};

    	\foreach \s in {0,...,5}
    	{%
    		\draw ( {-\s} , 0 ) node {$\times$};
    	}	
		\draw ( 2 , 0 ) node {$\times$};
    \end{tikzpicture}
 
 \tabularnewline
\end{tabular}

\caption{\label{fig:figure1} \emph{(Left)} Poles of the argument of the Cahlen-Mellin integral \eqref{eq:CahenMellinIntegral}. \emph{(Right)} Path in the complex plane over which the contour integration (\ref{eq:IntegrandZeta}) is performed. The semicircle of radius $R \rightarrow \infty$ is closed on the left.}

\par\end{centering}

\end{figure}
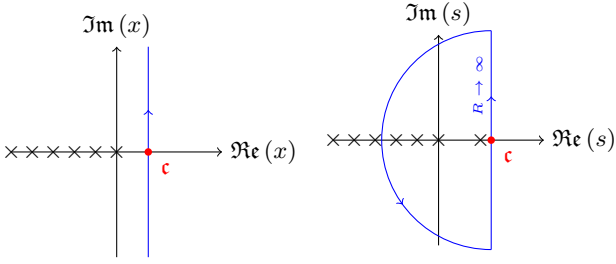

\begin{figure}[t]
\begin{centering}
\includegraphics[
scale=0.45,
keepaspectratio,trim={0 0.65cm 0 2.0cm},clip]{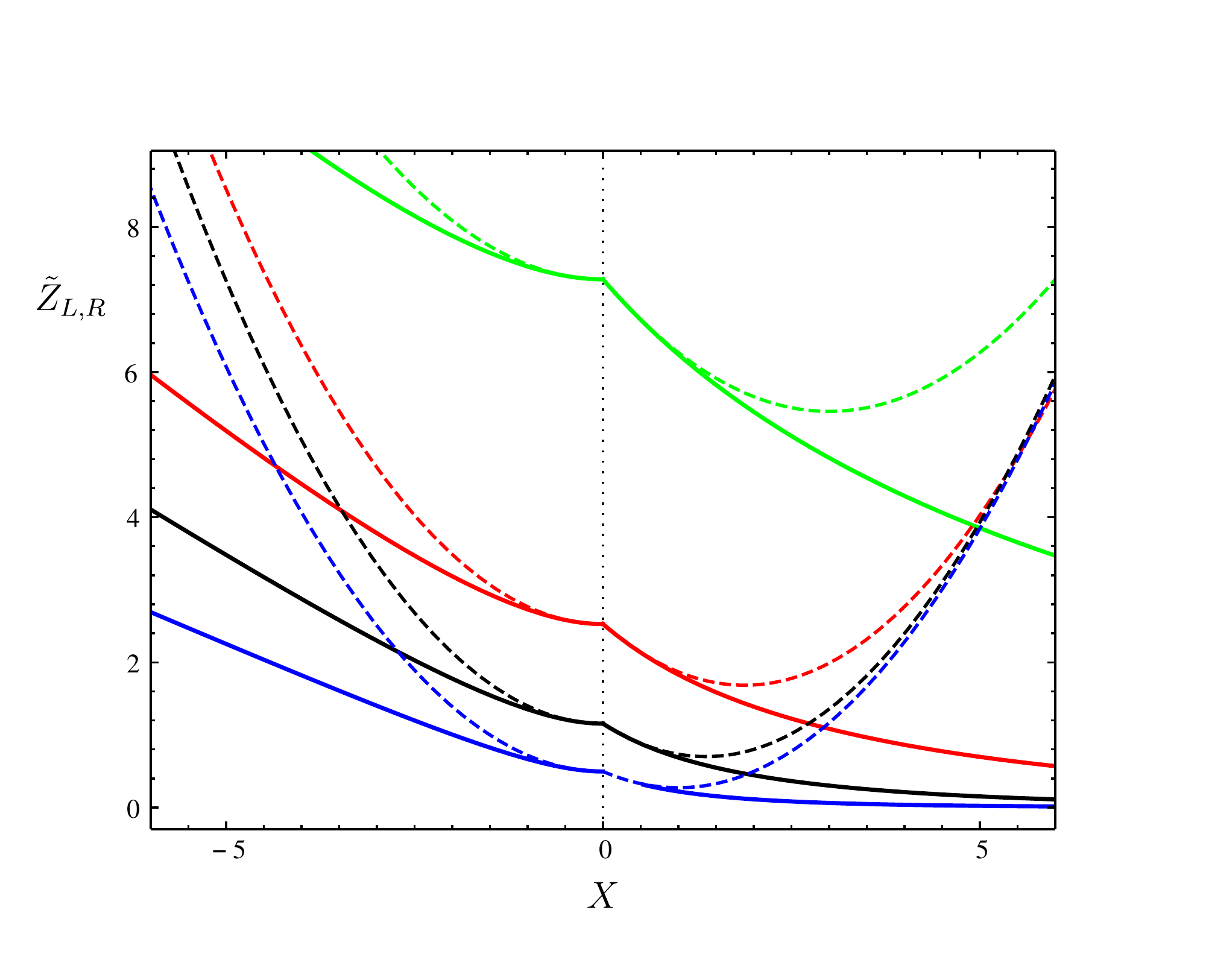}
\caption{Scaled partition function  for the commutative Dirac oscillator and its second order approximation at the critical point (dashed) as a function of the  scaled effective magnetic energy $X$, defined in \eqref{eq:xiLcomm},  for different values of the temperature:
$\tilde{\beta}=0.5, 0.8, 1.1, 1.5 $  respectively green, red, black and blue lines.  \label{fig:2} }
\end{centering}
\end{figure}

A different representation of the defining series in the  partition function that will reveal itself  
to be more effective when the phase switches from left to right, can be obtained using the Cahen-Mellin integral, as suggested for different systems in \cite{Brevik:2005ys} and
reviewed in \cite{Dariescu:2007,boumali2015one}.
The Cahen-Mellin integral is defined as
\cite{ParisKaminski200109} 
\begin{equation}
e^{-x}=\frac{1}{2\pi i}\int_{\mathfrak{c}-i\infty}^{\mathfrak{c}+i\infty}\Gamma\left(s\right)\: x^{-s}\ ds \quad \left(\left|\arg x\right|<\frac{1}{2}\pi;\, x\neq0\right)\label{eq:CahenMellinIntegral}
\end{equation}
 where $\mathfrak{c}$ is  real and $\mathfrak{c}>0$. The argument of the integral has poles for all $x=-n$, $n\in \mathbb{N}_0$ and the residuals at the negative poles are $\left(-1\right)^{n}/n!$ (see Fig. \ref{fig:figure1}).
Using (\ref{eq:CahenMellinIntegral}) to calculate the second term in the partition function (\ref{eq:ZSingleOscillator}) and including the sum into the integral, we have that the term
\begin{equation}
\xi_{L}\sum_{n_{r}=1}^{\infty}\ e^{-\beta mc^{2}\sqrt{1+\,\xi_{L}\, n_{r}}}
\end{equation}
can be rewritten as
\begin{equation}
\frac{\xi_{L}}{2\pi i}\int_{\mathfrak{c}-i\infty}^{\mathfrak{c}+i\infty} \frac{\Gamma\left(s\right)}{\left(\beta mc^{2}\right)^{s}}\:\sum_{n_{r}=1}^{\infty}\frac{1}{\left(1+\,\xi_{L}\, n_{r}\right)^{s/2}}\ ds. \label{eq:term2}
\end{equation}
The series in (\ref{eq:term2})  can be recognized as the series representation of the Hurwitz-$\zeta$ function $\zeta\left(v,a\right)=\sum_{n=0}^{\infty}\frac{1}{\left(n+a\right)^{v}}$ that converges 
only if $\ensuremath{\mathfrak{Re}\left(v\right)}>1$. Under this condition  the integral (\ref{eq:term2}) becomes
\begin{equation}
\frac{\xi_{L}}{2\pi i}\int_{\mathfrak{c}-i\infty}^{\mathfrak{c}+i\infty} \frac{\Gamma\left(s\right)}{\left(\beta mc^{2}\xi_{L}^{1/2}\right)^{s}}\:\:\zeta\left(\frac{s}{2},\frac{1}{\xi_{L}}+1\right)\ ds.  \label{eq:IntegrandZeta}
\end{equation}
 In contrast to what has been proposed in the recent literature on the topic (see \cite{boumali2015one} and references therein), we notice that the condition for the convergence of the series $s/2>1$ implies $\mathfrak{c} >2$.  The integral can therefore be evaluated with the method of residue once a proper closed path has been identified.
Because of the presence of the $\Gamma$-function that diverges as $\Gamma(s)=\left(2\pi\right)^{1/2}e^{-s}s^{s-\frac{1}{2}}\left[ 1-\mathcal{O}\left(s^{-1}\right)\right] $,
 the integration in \eqref{eq:IntegrandZeta} can be closed only on the left part of the complex plane, allowing to use the Cauchy's residue theorem on the poles $\left\{ s=2,0,\mathbb{Z}^{-}\right\} $ represented in Fig. \ref{fig:figure1}. Introducing the adimensional variable $\tilde{\beta}=\beta m\, c^2$, the partition functions for both the phases becomes
\begin{multline}
	Z_{L,R}=2\pi\left(\frac{\mathtt{L}mc}{h}\right)^{2}\times\\
	\times\left[\left(\frac{2}{\tilde{\beta}^{2}}-1\right)+\left(\frac{\Theta e^{-\tilde{\beta}}}{2}-\frac{1}{2}\right)\xi_{L,R}+\right.\\
	\left.+\sum_{n=1}^{\infty}\frac{\left(-\tilde{\beta}\right)^{n} \xi_{L,R}^{\frac{n}{2}+1}}{n!}\zeta\left(-\frac{n}{2},1+\frac{1}{\xi_{L,R}}\right)\right],
\label{eq:Zzeta}
\end{multline}
where now we use the subscript $L,R$ to indicate both the left and right phase and $\Theta$ is a step function that is equal to zero (one) for the right phase (left phase). 
This series representation can be seen as a series expansion in power of  $\beta$, namely a high-temperature power expansion.
From now on the quantity in the square brackets of \eqref{eq:Zzeta} will be called $\tilde{Z}_{L,R}$
\begin{equation}
Z_{L,R}=2\pi\left(\frac{\mathtt{L}mc}{h}\right)^{2} \tilde{Z}_{L,R}.
\end{equation}
The series representations of the partition function $ \tilde{Z}_{L,R}$ in terms of the $\zeta$-function  and  the Boltzmann representation introduced in \eqref{eq:ZSingleOscillator} can be both  evaluated numerically considering only a finite number of terms in the series. However, the Boltzmann representation encounters obvious numerical issues for small values of $\xi_{L,R}$, i.e. close to the value where the phase transition happens. These issues do not affect the  $\zeta$-function representation of the partition function and, as we will see in the next subsection, the partition function can be evaluated analytically for small values of  $\xi_{L,R}$.
 Fig. \ref{fig:2} shows $ \tilde{Z}_{L,R}$ in the $\zeta$-function representation keeping $n_{\text{max}}=100$ terms of the series in \eqref{eq:Zzeta}.
\subsection{Partition function near the critical point} \label{sec:CriticalPoint}
The partition function \eqref{eq:Zzeta} can be used to calculate the asymptotic expansion near the critical point, namely when $\xi_{L,R}\rightarrow 0$. This allows to analytically explore the regimes close to the phase transition and study the interplay between thermal  and quantum fluctuations. 
The asymptotic expansion of the Hurwitz-$\zeta$ function \cite{MR2124202,magnus2013formulas} is 
\begin{equation}
\zeta\left(s,a\right)=\frac{a^{1-s}}{s-1}+\frac{1}{2}a^{-s}+\frac{\mathtt{Z}\left(s,a\right)}{\Gamma\left(s\right)}\label{eq:ZetaExp}
\end{equation}
where the large-$a$ (Poincar\'e) asymptotic expansion of
 the function $\mathtt{Z}\left(s,a\right)$
\begin{equation}
\mathtt{Z}\left(s,a\right)\sim\sum_{k=1}^{\infty}\frac{B_{2k}}{\left(2k\right)!}\frac{\Gamma\left(2k+s-1\right)}{a^{2k+s-1}},\qquad\left|a\right|\rightarrow\infty \label{eq:ZetaExp2}
\end{equation}
is valid in $|\text{arg}\;\; a| < \pi$ and $B_{2k}$ denote the even-order Bernoulli numbers. 
Writing explicitly the sum \eqref{eq:ZetaExp2} in  the partition function $\tilde{Z}_{L,R}$  gives
\begin{multline}
\tilde{Z}_{L,R}=\frac{2}{\tilde{\beta}^{2}}-1-\left(1-\Theta e^{-\tilde{\beta}}\right)\frac{\xi_{L,R}}{2}+\\
+\sum_{n=1}^{\infty}\frac{\left(-\tilde{\beta}\right)^{n}}{n!}\left\{ \left[\frac{1}{2}\xi_{L,R}-\frac{2\left(\xi_{L,R}+1\right)}{n+2}\right]\left(\xi_{L,R}+1\right)^{\frac{n}{2}}+\right.\\
\left.+\sum_{k=1}^{\infty}\frac{B_{2k}}{\left(2k\right)!}\frac{\Gamma\left(2k-\frac{n}{2}-1\right)\:}{\Gamma\left(-\frac{n}{2}\right)}\frac{\xi_{L,R}^{2k}}{\left(\xi_{L,R}+1\right)^{2k-\frac{n}{2}-1}}\right\} . \label{eq:ZtildPreBinomial}
\end{multline}
Using the binomial series and the relation between Euler gamma functions
\begin{equation}
\Gamma\left( -\frac{n}{2} \right) \; \Gamma\left( 1 + \frac{n}{2} \right) = \frac{\pi}{\sin\left( - \pi n/2 \right)},
\end{equation}
one can rewrite the terms in the curly brackets in (\ref{eq:ZtildPreBinomial}) as
\begin{multline}
 \sum_{w=0}^{\infty} \frac{1}{w!} \left\{ \left[\frac{1}{2}\xi_{L,R}-\frac{2\left(\xi_{L,R}+1\right)}{n+2}\right] 
 \frac{\Gamma \left(1+ \frac{n}{2} \right)}{\Gamma \left(1+ \frac{n}{2} - w \right)} \xi_{L,R}^{w} +
 \right.\\
\left.+\sum_{k=1}^{\infty}\frac{B_{2k}}{\left(2k\right)!}\frac{\Gamma\left( 1+ \frac{n}{2} \right)\:}{\Gamma\left( -2k +\frac{n}{2}+2-w \right)} \xi_{L,R}^{2k+w} \right\}
\end{multline}
defining in this way the partition function at all orders  in $\xi_{L,R}$ near the critical point.
At the second order in $\xi_{L,R}$, one can sum the partition function at all orders in $\beta$, so that, near the critical point,  it reads
\begin{multline}
\tilde{Z}_{L,R} =\frac{2e^{-\tilde{\beta}}}{\tilde{\beta}}\left[1+\frac{1}{\tilde{\beta}}+\left(\Theta-1\right)\frac{\tilde{\beta} \xi_{L,R}}{4}+\frac{\tilde{\beta}^{2}\,\xi_{L,R}^{2}}{48}\right] + 
\\ 
+{\cal O}\left(\xi_{L,R}^{3}\right).\label{eq:Z2ndOrder}
\end{multline}
This expression  is a continuous but non-analytic function in $\xi_{L,R}$ and, in fact,  $\Theta$ brings a discontinuity in the first order derivatives. The function to the second order \eqref{eq:Z2ndOrder} is compared with \eqref{eq:Zzeta} in Fig. \ref{fig:2}.

\section{Magnetization at finite temperature} \label{sec:magnetiz}

At this point, it is possible to use the derived expression for the partition function \eqref{eq:Zzeta} to calculate a physical observable quantity like the magnetization~\cite{Yoshioka1992,Andersen1995}. Indeed, in \cite{Panella:2014hga} the magnetization has been proposed as the quantity able to distinguish between the phases of the system. 

\begin{figure}[t]
\begin{centering}
\includegraphics[
scale=0.45, 
keepaspectratio,
trim={0 1.1cm 0 2.5cm},clip
]{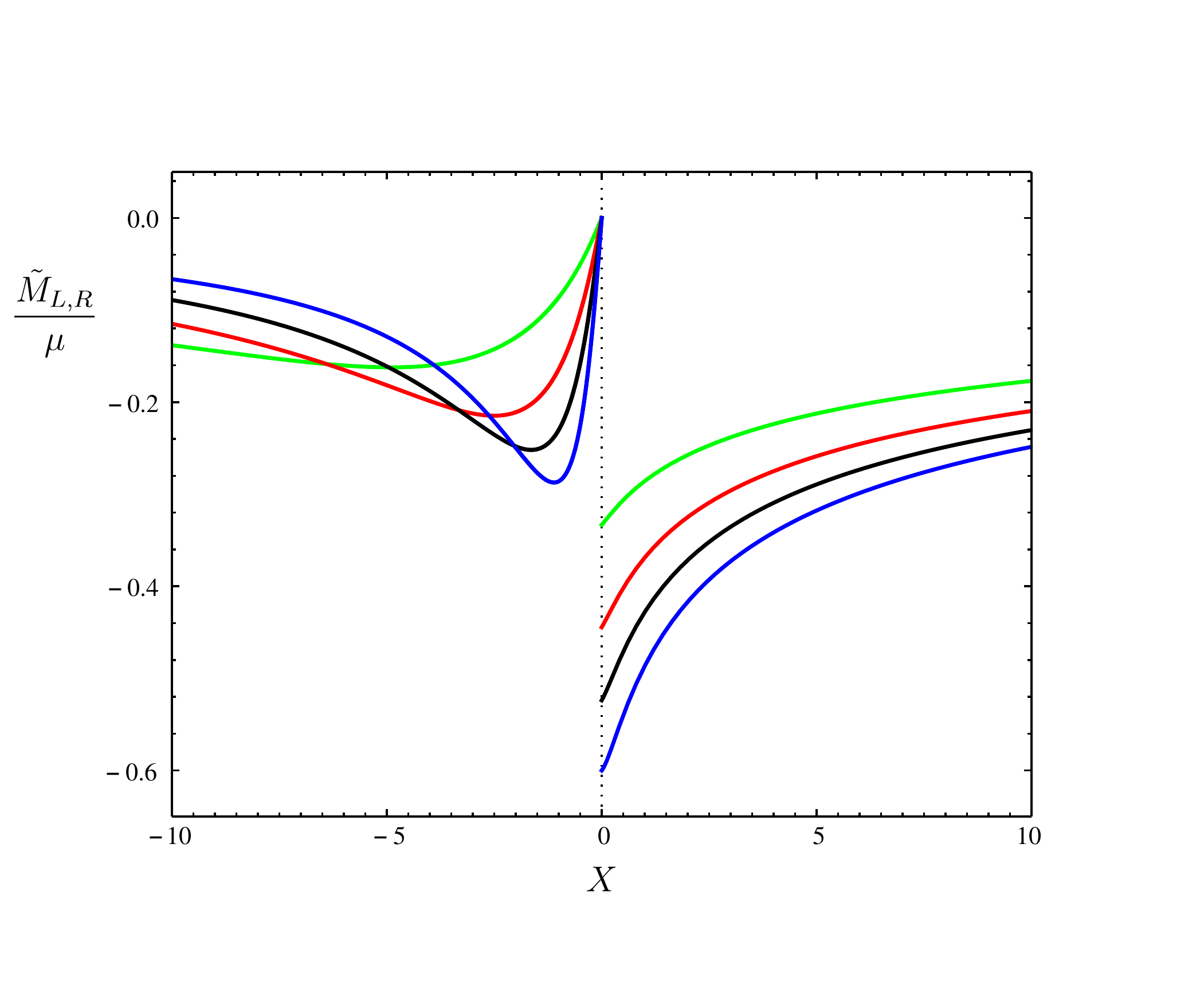}
\caption{Scaled magnetization of the  commutative Dirac oscillator as function of the  scaled effective magnetic energy  $X$, defined in \eqref{eq:xiLcomm},
for different values of the temperature $\tilde{\beta}=0.5, 0.8, 1.1, 1.5  $  respectively  green,  red, black and blue lines. 
}\label{fig:MagnComm}
\end{centering}
\end{figure}

The magnetization for a system at finite temperature, can be defined as \cite{Kubo1990} 
\begin{gather}
M=k_{B}T\frac{\partial\log Z}{\partial B}=\frac{1}{\beta\,Z}\frac{\partial Z}{\partial\xi}\frac{\partial\xi}{\partial X}\frac{\partial X}{\partial B}.\label{eq:Mformula}
\end{gather}
In order evaluate the derivative $\partial Z/\partial\xi_{L.R}$ one can use in \eqref{eq:Zzeta} the following identity \cite{NIST:DLMF} 
\begin{gather}
\frac{\partial\zeta\left(s,a\right)}{\partial a}=-s\:\zeta\left(s+1,a\right)\ \mbox{for }s\neq0,1\,\mathfrak{Re}a>0. 
\end{gather}
The magnetization evaluated numerically with $n_\text{max}=60$ is illustrated in Fig. \ref{fig:MagnComm}. 
Interestingly, we find that the magnetization in the left phase, has opposite sign with respect to the single state result obtained at $T=0$ \cite{Panella:2014hga}. At finite temperature, in $\xi_L =\xi_R = 0$ the magnetization manifests a discontinuity.

Analogously to what has been done for the partition function in the previous sections, we now inspect the magnetization behaviour at values of the magnetic field close to the critical values. 
To approximate the full partition function in the neighborhood of $\xi_L=\xi_R=0$, we can either follow the steps described in subsection \ref{sec:CriticalPoint} for the full partition function resulting of \eqref{eq:Mformula}, or we can use  \eqref{eq:Z2ndOrder} in   \eqref{eq:Mformula}. 

At the critical point, the magnetization is not a continuous function of the magnetic field and can be written as a series expansion around $X=0$ at all order in $\beta$ as 
\begin{multline}
\frac{M_{R,L}}{\mu}= 
-\frac{\tilde{\beta}}{\tilde{\beta}+1}H\left(X\right)+\\
 +\left(\frac{\tilde{\beta}}{\tilde{\beta}+1}\right)^{2}\left[\frac{2}{3}+\tilde{\beta}\left(H\left(-X\right)-\frac{1}{3}\right)\right]\,X  +
 \mathcal{O}\left(X^{2}\right)
 ,
\end{multline}
where 
$H(x) $ is the Heaviside step function. The discontinuity with respect to the magnetic field at the critical point $X=0$ is
\begin{equation}
    |\Delta M_{\text{cr}}| = 
    \mu\, \frac{\, \tilde{\beta} }{ (1+\tilde{\beta} )}. 
    \label{eq:Mgap}
\end{equation}
In the limit $\tilde{\beta} \rightarrow 0 $, the thermal fluctuations overcome the quantum ones, the gap disappears and the magnetization is zero for both the chiralities. 

\section{Other thermodynamic functions}
\begin{figure*}[t]
	\begin{centering}
\includegraphics[height=12cm,trim={0 1.1cm 0 2.5cm},clip]{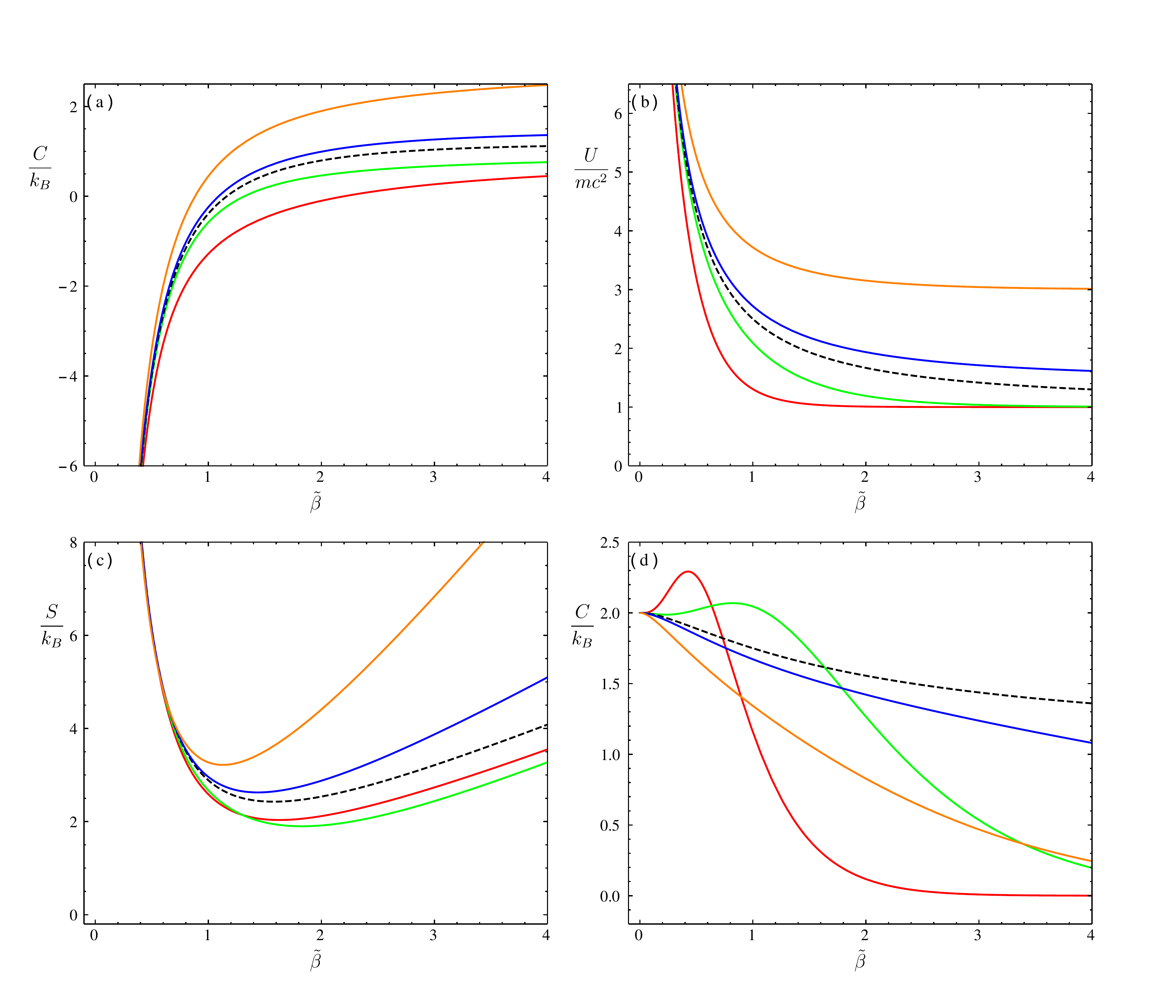}\end{centering}
		\caption{Various thermodynamic functions versus the scaled inverse temperature $\tilde{\beta}=mc^2 / k_B T$ for different values of the magnetic field, corresponding to  $X=-4.5, -1.3, 0, 0.3,2 $   respectively red, green, dashed-black, blue, orange. The  $X=0$ case (dashed black) represents the critical value of the magnetic field. Upper left panel (a): the free energy $F/mc^2$; upper right panel (b): the internal energy $U/mc^2$; lower left (c): the entropy $S/k_B$ and finally in the lower right panel (d) the specific heat $C/k_B$.  In the last one, it is visible the difference between  the red and green (left phase) and the blue and orange (right phase).}  \label{fig:Free}
	
\end{figure*}
Once the partition function has been calculated, the thermodynamics
of the system can be fully explored. Using
\begin{equation}
\begin{aligned}
F=-\frac{1}{\beta}\ln Z,\qquad U=-\frac{\partial}{\partial\beta}\ln Z,\\
S=\, k_{B}\beta^{2}\frac{\partial F}{\partial\beta},\qquad C=-k_{B}\beta^{2}\frac{\partial U}{\partial\beta},
\end{aligned}\label{eq:ThermQuant}
\end{equation}
in Fig.~\ref{fig:Free}, we plot the thermodynamic quantities for the commutative system. The specific heat as function of the $\beta$  manifests a non-monotonic behaviour in the left phase, while it becomes monotonic in the right phase.

\section{Noncommutative Dirac Oscillator with a constant magnetic field} \label{sec:NCoscillator}

In the previous sections we described the Dirac oscillator under a uniform magnetic field and its associated QPT in a  $2D$ space-time.
Now, we move to the noncommutative system, indeed,
studies in quantum gravity and string theory propose that the space-time can be fundamentally or effectively described by a generalization of geometry  where, locally, coordinates do not necessarily commute \cite{Bertolami2005,Gamboa2002,Ito1967,Bellucci2001b,Duval2000,Gamboa2001a,Gamboa2001,Horvathy:2010wv,Nair2001,Smailagic2002,Smailagic2002a}.
Moreover, recent developments in condensed matter physics, suggest the use of the noncommutative geometry framework to encode geometrical properties of topological quantum systems \cite{2012PhRvB..86c5125N,Haldane:2011ia,2012PhRvB..85x1308P,2012PhRvB..85g5128B}. 

The study of the  Dirac oscillator in the presence of noncommutative coordinates and momenta and a constant magnetic field in a $2$-dimensional space at zero-temperature has been carried out in \cite{Panella:2014hga}. The system that in the commutative case undergoes a quantum phase transition at finite magnetic field, also exhibits in the noncommutative case another phase transition at a higher magnetic field.  The third region appears to have the same chirality of the first one providing a re-entrant quantum phase transition \cite{2016:PanellaRoyReE}. 

In this section, we will study the noncommutative system at finite temperature, the parametrization that has been used for the exact calculation of the spectrum in \cite{Panella:2014hga} will allow us to exploit all the calculation done in the previous sections and easily study the QPTs in the noncommutative space.

\subsection{Energy levels}

The Hamiltonian for the $(2 + 1)$-dimensional Dirac oscillator in the noncommutative
plane with a homogeneous magnetic field can be written in a way similar to (\ref{eq:hamiltonian})
\begin{equation}
  \hat{H}=
  c\bm{\sigma}.
  (\hat{\bm{p}}-im\omega\sigma_z\hat{\bm{x}}+\frac{e}{c}\,\hat{\bm{A}})
  +\sigma_{z}\,mc^{2}\,,\label{h1}
\end{equation}
where now the \emph{hat} indicates the noncommutative operators.
In this framework, the commutation relation between coordinates and momenta are given by \cite{Bertolami2005}
\begin{equation}
  [{\hat{x}},{\hat{y}}]=i\theta,
  ~~~~[{\hat{p}}_{x},{\hat{p}}_{y}]=i\eta,
  ~~~~[{\hat{x}}_{i},{\hat{p}}_{j}]=i\hbar(1+\frac{\theta\eta}{4\hbar^{2}})\delta_{ij},
    \label{NCEQ}
\end{equation}
where $\theta,\eta\in\mathbb{R}$. 
The noncommuting coordinates and momenta can be expressed in terms of
commuting ones using the Seiberg-Witten map and are given by
\begin{equation}
\begin{aligned}
{\hat x}=\displaystyle {x-\frac{\theta}{2\hbar}p_y,~~~~{\hat p}_x=p_x+\frac{\eta}{2\hbar}y}\,,\\
{\hat y}=\displaystyle{y+\frac{\theta}{2\hbar}p_x,~~~~{\hat p}_y=p_y-\frac{\eta}{2\hbar}x}\label{rel}\,.
\end{aligned}
\end{equation}
The Hamiltonian (\ref{h1}) can be rewritten as 
\begin{equation}
  H=
  c\left(\begin{array}{cc}
           mc & {\hat{\Pi}}_{-}\\
           {\hat{\Pi}}_{+} & -mc
         \end{array}
  \right),\label{h2}
\end{equation}
where $\hat{\Pi}_{\pm}$ are given by 
\begin{multline}
{\hat\Pi}_{\pm} = \left( 1-
 \frac{\tilde{\omega}-\omega}{\omega_\theta}\right)(p_x \pm i\, p_y)\,\pm \\ \pm\, i\, m\, (\tilde{\omega} -\omega -\omega_\eta) (x \pm i y)
\end{multline}
after introducing the following frequencies out of the parameters of the system: 
\begin{equation}
\tilde{\omega}=\frac{eB}{2mc},\,\qquad \omega_\theta = \frac{2\,\hbar}{m\theta},\, \qquad \omega_\eta=\frac{\eta}{2\hbar m}.
\end{equation}
With this paramatrization it is clear that there are two critical values of the magnetic field 
\begin{gather}
\label{bc}
B_{\text{cr}}=\frac{2 mc}{e} (\omega +\omega_\eta)=\frac{2c}{e}(m\omega+\frac{2\eta}{\hbar}) \,,\\
\label{bcrit*}
B_{\text{cr}}^*=\frac{2 mc}{e} (\omega +\omega_\theta)=\frac{2c}{e}(m\omega+\frac{\hbar}{\theta}) \, .
\end{gather}
such that for both $B=B_{\text{cr}}$ and $B=B_{\text{cr}}^*$,    there are no interactions in the model and the Hamiltonian represents a free particle (only kinetic energy). Since the critical value $B_{\text{cr}}$ depends on the momentum noncommutative parameter $\eta$, this shifts the value of the critical field relative to the value in \eqref{eq:BcrComm}.The space noncommutativity ($\theta \ne 0$) instead introduces the additional critical value for the magnetic field ($B_{\text{cr}}^*$) \cite{Panella:2014hga}.

Interestingly, the parameters used in the noncommutative massive Dirac oscillator can be re-assorbed in such a way that one can study the single phases independently. In fact, the behaviour of the single phases can be described
regardless of the parameters encoding the noncommutative physics \cite{Panella:2014hga}.
Therefore, although the noncommutative system has two more parameters,  
they can recast into the parameter $\xi_{L,R}$ allowing us to use most of the calculations performed in the previous sections for the commutative case. 
The main difference now is that the magnetic field and the noncommutative parameters are related  by  
\begin{eqnarray}
    \xi_{L}&=&-4\frac{mc^{2}}{\omega_{\theta}\hbar}\left(\frac{\omega_{\theta}\hbar}{mc^{2}}-X\right)\left(X-\frac{\omega_{\eta}\hbar}{mc^{2}}\right) \label{eq:xiL} 
\end{eqnarray}  
while it is still valid the relation  $\xi_R = -\xi_L$. Phase changes occur at critical values of the magnetic field that makes the parameter $\xi_{L,R}$ null. Note that, differently from the commutative case \eqref{eq:xiLcomm}, in the noncommutative case the equation \eqref{eq:xiL} is quadratic in $B$  and consequently it admits two critical points.  
\begin{figure}[t]
	\begin{centering}
		\includegraphics[
		trim={0 0.7cm 0 2.4cm},clip,
		scale=0.45
		]{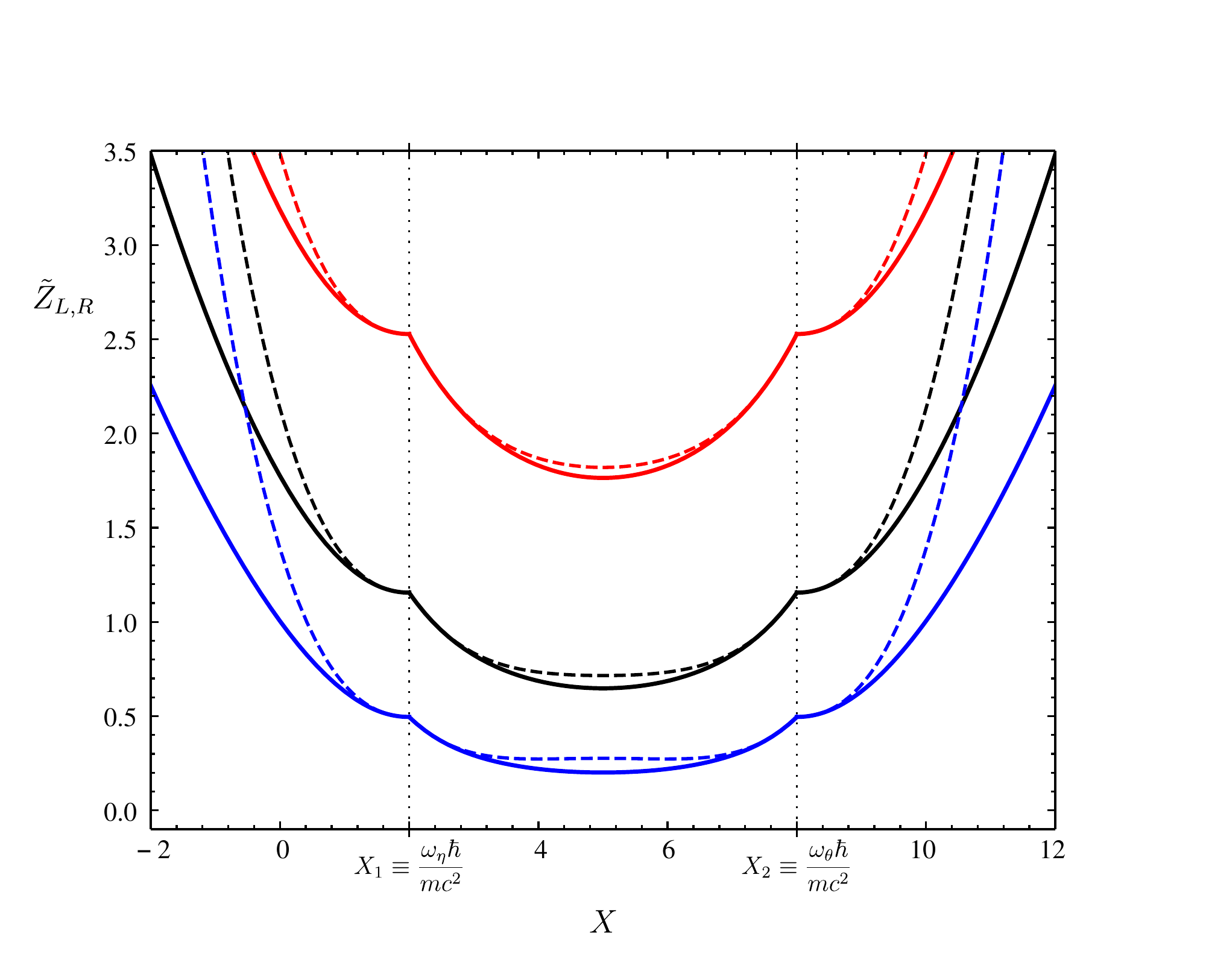}
		
		\caption{Scaled partition function in the noncommutative case and its approximation near the critical points $X_1$, $X_2$ (dashed) with
			$X_1={\omega_{\eta}\hbar}/{mc^{2}}=2$, 
			$X_2=\left({\omega_{\theta}\hbar}/{mc^{2}}\right)=8$ and
			$\tilde{\beta}= 0.8, 1.1, 1.5 $ for respectively red, black  and blue lines.\label{fig:ZNonComm}}
	\end{centering}
\end{figure}
\subsection{Energy levels and partition function}
In Section \ref{sec:energy-levels} we reviewed the spectrum of the system as a function of the parameter $\xi_{L,R}$. As mentioned above, for the noncommutative case, the spectrum remains equal to \eqref{spectrumL1} and  \eqref{spectrumR} as a function of the parameter $\xi_{L,R}$ that now becomes \eqref{eq:xiL}, quadratic in $B$.   
Since the density of the energy states is also a function of the parameters $\xi_{L,R}$, its contribution to the partition function of each energy level in the noncommutative case coincides with the one in the commutative case \eqref{eq:StateDensity}. The resulting partition function is illustrated in Fig. \ref{fig:ZNonComm}. In the plot is also represented the outcome of the approximation once the constant $\xi_{L,R}$ in \eqref{eq:xiL} is replaced in  \eqref{eq:Z2ndOrder}.
The partition function can be used to calculate the other thermodynamic quantities using \eqref{eq:ThermQuant}. In particular, the heat capacity is shown in Fig. \ref{fig:CNNcom}. 
    In \cite{boumali2013thermal} a different noncommutative model has been studied without considering the degeneracy of the states. 

\subsection{Magnetization}
\begin{figure}
\begin{centering}
\includegraphics[
trim={0 0.7cm 0 2.4cm},clip,
scale=0.45
]{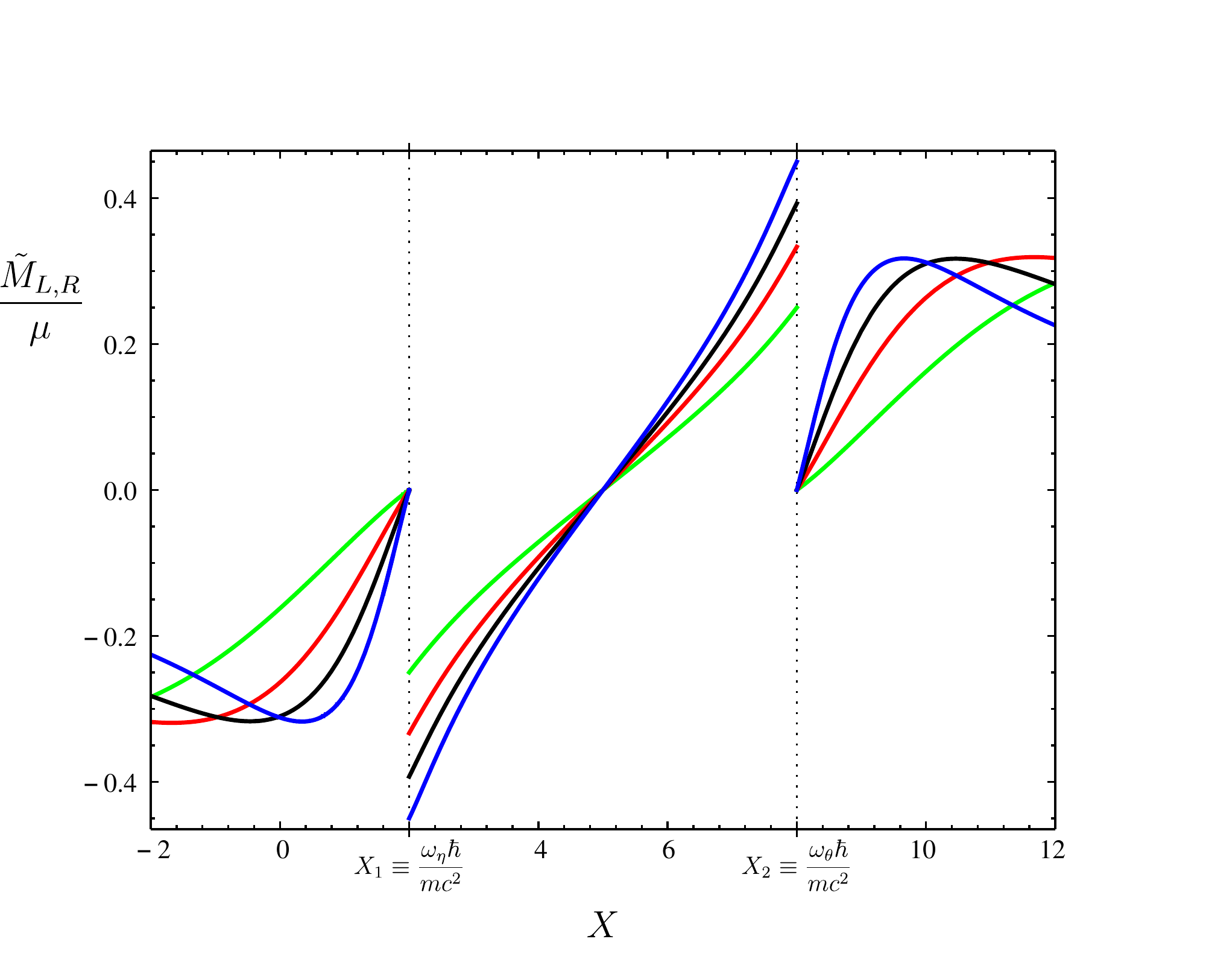}

\caption{Scaled magnetization  in the noncommutative case as function of the  scaled effective magnetic energy  $X$, defined in \eqref{eq:xiLcomm},  for different values of the rescaled inverse temperature $\tilde{\beta}=0.5, 0.8, 1.1, 1.5  $  respectively  green,red, black and blue lines. The parameters are fixed as in Fig. \ref{fig:ZNonComm}.} \label{fig:Magn}
\end{centering}
\end{figure}

The magnetization for the noncommutative Dirac oscillator case is illustrated in Fig. \ref{fig:Magn}.  The figure shows the presence of a fixed point at $M=0$ for every $T$. The point where the magnetization changes its sign appears only in the noncommutative case and is a consequence of the term $ \partial \xi / \partial X $ in \eqref{eq:Mformula}, that is a non constant function of $X$:
\begin{multline}
	\frac{\partial\xi_{L,R}}{\partial X}=\pm
	4\frac{mc^{2}}{\omega_{\theta}\hbar}\left[2\, X-\frac{\hbar\left(\omega_{\eta}+\omega_{\theta}\right)}{mc^{2}}\right]= \\
	 =\pm 4\sqrt{\frac{m\,c^{2}}{\hbar\,\omega_{\theta}}
	 \xi_{L,R} \pm 
	 \frac{(\omega_{\eta}-\omega_{\theta})^{2}}{\omega_{\theta}{}^{2}}} \label{eq:Derxix}
\end{multline} 
where the plus sign is for the $L$ phase and the minus sign is for the $R$ phase. From the first line of \eqref{eq:Derxix}, we find that
 at the point $X_0 = \frac{\hbar  (\omega_\eta +\omega_\theta )}{2 mc^2}$ the magnetization change{\bf s} sign. Therefore, the point where the magnetization smoothly vanishes is the midpoint between the two critical points. This feature of the noncommutative system is independent  of the temperature and was also present in the analysis at $T=0$ \cite{Panella:2014hga}.   
Near the critical point, we can evaluate the magnetization in $\xi_{L,R}$ using the second line of \eqref{eq:Derxix} and \eqref{eq:Z2ndOrder} in \eqref{eq:Mformula}, obtaining:   
\begin{gather}
\frac{\partial\xi_{L,R}}{\partial X} = \pm\,4\,\left[\left|\frac{(\omega_{\eta}-\omega_{\theta})}{\omega_{\theta}}\right|\pm\frac{1}{2}\frac{m\,c^{2}}{\hbar\,}\frac{\xi_{L,R}}{\left|\omega_{\eta}-\omega_{\theta}\right|}+\mathcal{O}(\xi^{2})\right],
\end{gather}
and the other terms in \eqref{eq:Mformula} read
\begin{multline}
\frac{1}{\tilde{\beta}}\frac{1}{Z}\frac{\partial Z}{\partial\xi} =
\frac{\tilde{\beta}}{4(\tilde{\beta}+1)}\left(\Theta-1\right)+ \\ +
 \left[\frac{\tilde{\beta}}{4(\tilde{\beta}+1)}\right]^{2}\left[\frac{2}{3}+\tilde{\beta}\left(\Theta-\frac{1}{3}\right)\right]\xi+
\mathcal{O}\left(\xi^{2}\right).
\end{multline}
Then the discontinuity of the magnetization at the two critical points, for the noncommutative case is
\begin{gather}
\Delta M=\pm
\mu\, \left|\frac{(\omega_{\eta}-\omega_{\theta})}{\omega_{\theta}}\right|\frac{\tilde{\beta}}{(\tilde{\beta}+1)},
\end{gather}
and we can notice that if $\eta$, the noncommutative parameter related with the momenta,  vanishes
then the transitions become independent from $\theta$, the coordinates noncommutative parameter.

\section{Conclusions}\label{sec:conclusions}

Over the years, the Dirac oscillator has become a valuable tool for various branches of physics, mostly because it is one of the few relativistic systems whose exact solutions are known. 
The $(2+1)$-dimensional case, namely when an electron is constrained to live in a plane, manifests an intrinsic chiral behavior of its states. An external magnetic field interacts at the quantum level with the chirality properties of the system and, if strong enough, it forces an abrupt switch in the electron chirality. This behavior studied at zero temperature,  distinguish two phases of the electron at different regimes, and their change is recognized to be a quantum phase transition \cite{2008PhRvA77:Delgado}. Magnetization shows a discontinuity at a finite value of the magnetic field \cite{Panella:2014hga}. 
The quantum nature of this behavior raises the question on how thermal fluctuations interacts with the system and when the thermal disorder destroys these quantum effects. 
In this paper, we analyze the interplay between these two physical phenomena. 
In particular, we quantify the discontinuity of the magnetization of the QPT at finite temperature  $\Delta M \propto \hbar \left[ \frac{1}{k_B\,T}+O(\frac{1}{k_B\,T})^2\right]$. This discontinuity is a quantum phenomenon that tends to zero for high temperatures.
While the critical values of the system are not affected by changes in temperature, we noticed a crucial difference from the system at $T=0$ studied in \cite{Panella:2014hga}: the magnetization at finite temperature has regimes with opposite sign with respect to the single state (zero-temperature) case. This opens to the possibility of a second phase transition at low temperatures, where the statistics of the states becomes quantum and cannot be described by Boltzmann statistics. In Fig. \ref{fig:Free} 
we report various thermodynamic quantities that can be evaluated once the partition function is provided. The specific heat shows different behaviors at different phases of the system. 

The series representations of the partition function proposed in this paper allow to explore the system in diverse regimes: far, close and at the critical points. The representation in terms of the Hurwitz-zeta function  fixes some misunderstandings present in the literature regarding which poles play a role in the evaluation of the partition function. 

We also generalize our calculations to the Dirac oscillator in noncommutative momenta and coordinates. The interest for this system comes from the fact that at zero temperature it manifests a re-entrant quantum phase transition, that we find is present also at finite temperature. 
Interestingly, there are systems in condensed matter physics that in certain regimes appear to have behavior similar and some times mathematically equivalent to systems living in a noncommutative space-time  \cite{2012PhRvB..86c5125N,Haldane:2011ia,2012PhRvB..85x1308P,2012PhRvB..85g5128B}.   
This opens the avenue of an \emph{analogue noncommutativity} that is the possibility of studying, also experimentally, systems that are mathematically equivalent to those with noncommutating coordinates and/or momenta but are condensed matter systems living in an ordinary commutative space.
The ``analogue'' approach to theoretical models that are too extreme or too weak to be measured directly has already been successfully employed in high-energy physics, and in particular in gravity with the so called analogue gravity \cite{Barcelo:2005fc}. This program is in constant development and already lead to important successes such as the  measures of the analogue Hawking radiation \cite{Weinfurtner:2010nu, Belgiorno:2010wn, Steinhauer:2014dra,Steinhauer:2015saa}, phenomenon  that seemed  to be  only a theoretical model. 
It is interesting and challenging to find an analogue system that would allow the experimental measure of the re-entrant phase transition described in this paper, and we leave it for future investigation.

\section*{Acknowledgements}
A.M.F. would like to thank the Stiftung Giersch for the generous support during this work. D.M. acknowledges L. Malag\`o for discussions and support.

\medskip

\bibliography{DiracOscillator}
\end{document}